\documentclass[preprint,showpacs,floats,nofootinbib,superscriptaddress,color,
  showkeys]{revtex4}

\usepackage{bm}
\usepackage{amsmath}
\usepackage{graphicx}
\usepackage{graphics}
\usepackage[dvips]{color}
\newcommand{\be}{\begin{equation}} \newcommand{\ee}{\end{equation}}
\newcommand{\ba}{\begin{eqnarray}} \newcommand{\ea}{\end{eqnarray}}
  \newcommand{\nk}{{\bf k}} \newcommand{\nq}{{\bf q}}
\newcommand{\np}{{\bf p}} 
 \newcommand{\nkappa}{\mbox{\boldmath
    $\kappa$}}

\begin{document}
\title{Superscaling and neutral current quasielastic
neutrino-nucleus scattering}
\author{J.E. Amaro}
\affiliation{Departamento de F\'\i sica Moderna,
Universidad de Granada,
18071 Granada, SPAIN }
\author{M.B. Barbaro}
\affiliation{Dipartimento di Fisica Teorica, Universit\`a di
Torino and INFN, Sezione di Torino, Via P. Giuria 1, 10125 Torino,
ITALY }
\author{J.A. Caballero}
\affiliation{Departamento de F\'\i sica At\'omica, Molecular y Nuclear,
Universidad de Sevilla, Apdo. 1065, 41080 Sevilla, SPAIN }
\author{T.W. Donnelly}
\affiliation{Center for Theoretical Physics, Laboratory for Nuclear Science
and Department of Physics,
Massachusetts Institute of Technology,
Cambridge, MA 02139, USA }

\date{\today}

\begin{abstract}

The superscaling approach is applied to studies of neutral current
neutrino reactions in the quasielastic regime. Using input from
scaling analyses of electron scattering data, predictions for
high-energy neutrino and antineutrino cross sections are given and
compared with results obtained using the relativistic Fermi gas
model. The influence of strangeness content inside the nucleons in
the nucleus is also explored.

\end{abstract}



\maketitle

\section{Introduction}
\label{sec:intro}

Inclusive electron scattering at intermediate to high energies
from nuclei is known to exhibit the phenomenon of scaling and
superscaling~\cite{Alberico:1988bv,Day:1990mf,Donnelly:1998xg,Donnelly:1999sw,
Maieron:2001it,Barbaro:2003ie,Barbaro:1998gu}. At sufficiently
high energies, typically at least 500 MeV, one sees that near the
quasielastic peak the cross section may be analyzed in terms of a
reduced response obtained by division by a suitable N- and
Z-weighted single-nucleon electromagnetic cross section and
plotted against an appropriate kinematic variable to see the
scaling behaviour. First, when the reduced cross section is seen
to depend only on this kinematic variable --- the scaling variable
--- and not on the momentum transfer one has scaling of the first
kind. Second, if the reduced cross section and scaling variable
have been made dimensionless via removal of the momentum scale
characteristic of a given nucleus, and the results are seen to be
independent of the particular nuclear species, one has scaling of
the second kind. When both types of scaling behaviour occur one
says that the cross sections exhibit superscaling. In the
above-cited studies the appropriate reduced cross sections and
scaling variables have been discussed in depth.

One finds that in the relevant energy range in the region below
the quasielastic (QE) peak, usually called the scaling region,
scaling of the second kind is found to be excellent and scaling of
the first kind to be quite good. Above the peak scaling of the
second kind is good; however, scaling of the first kind is clearly
violated. The last occurs for well-understood reasons, namely, in
that region one has processes other than quasi-free knockout of
nucleons playing an important role. Specifically, the most obvious
reaction mechanism is that of exciting a nucleon in the nucleus to
a delta, which subsequently decays into a nucleon and a pion.
Since the elementary cross section for that process is not the
elastic eN cross section used in defining the scaling function
introduced above, and since the scaling variable used in the usual
analysis assumes the kinematics of the elastic process $N \to N$,
rather than of $N \to \Delta$ which would now be appropriate, it
is not surprising that scale-breaking occurs. Additionally, meson
exchange current effects are known to violate the scaling
behaviour, although from modeling in this high-energy
regime~\cite{Amaro02,Amaro:2001xz,Amaro03,Amaro98,DePace03,DePace04}
their effects appear not to be the dominant ones.

What was appreciated for the first time in recent
work~\cite{Amaro:2004bs} is that it is possible to pursue an
approach where both the QE process is active (with its reduced
response and scaling variable) and {\em also} incorporate the
inelastic process in the $\Delta$-region (with its corresponding
reduced response and scaling variable). We shall roughly refer to
the region of excitation forming a peak that lies above the
maximum of the QE response as the ``$\Delta$-peak'', although it
should be understood that the modeling actually includes the full
inelastic response on a nucleon (resonant plus non-resonant) for
kinematics where the $\Delta$(1232) is dominant.\footnote{For
still higher-lying excitations and DIS a different approach must
be taken (see, for example, \cite{Barbaro:2003ie}).}
In~\cite{Amaro:2004bs} it was shown that an excellent
representation of the total inclusive electron scattering cross
section from the scaling region up to the peak of the
$\Delta$-region is attained by inverting the procedure. Using the
two scaling functions, one for QE scattering and one for the
$\Delta$-region, along with the corresponding N- and Z-weighted
elastic ($eN \to eN$) and inelastic ($eN \to e'\Delta$) electron
scattering cross sections one finds excellent agreement with
existing high-quality data over a wide range of kinematics and for
various nuclear species. Of considerable importance for what was
discussed in the rest of~\cite{Amaro:2004bs} and will be discussed
in the present work is the fact that the quality of the analysis
requires the phenomenological scaling functions to be quite
asymmetric, with relatively long tails extending in the direction
of higher energy loss (positive values of the scaling variables).
Such is not typically the case with most models, these almost
always being more nearly symmetrical about the peak in the scaling
function (see, however, \cite{Caballero:2005sj} where in at least
one case the correct behaviour has been obtained in a model). This
fact casts considerable doubt on most existing models for
high-energy scattering in the QE and $\Delta$ regimes if
high-quality results (say better than 25\%) are desired.

Having met with success in extending the scaling and superscaling
analyses from the scaling region, through the QE peak region and
into the $\Delta$ region, in~\cite{Amaro:2004bs} the scaling ideas
were inverted: given the scaling functions one can just as well
multiply by the elementary charge-changing (CC) neutrino cross
sections now to obtain the corresponding CC neutrino and
antineutrino cross sections on nuclei for intermediate to high
energies in the same region of excitation. Other related work is
presented in ~\cite{Caballero:2005sj,Amaro:2005dn}. Given the ability
of this scaling approach to reproduce the electron scattering
cross sections, in contrast to most direct modeling which fails in
detail to do so, we believe that such predictions for the
analogous CC neutrino reactions should be very robust. Clearly
such results are of relevance for on-going studies of neutrino
reactions and neutrino oscillations in this intermediate-energy
regime.

In the present study these scaling and superscaling ideas are
carried a step further to include neutral-current (NC) neutrino
and antineutrino scattering cross sections, in this case for
scattering from $^{12}$C. Specifically, the goal is to obtain
results using the same analysis as discussed above (and in detail
in~\cite{Amaro:2004bs}) for the reactions $^{12}$C($\nu,p)\nu$X,
$^{12}$C(${\bar{\nu}},p){\bar{\nu}}$X involving proton knockout
and $^{12}$C($\nu,n)\nu$X, $^{12}$C(${\bar{\nu}},n){\bar{\nu}}$X
involving neutron knockout in the QE regime, the $\Delta$-regime
being left for a subsequent study.

A new feature emerges with such a goal in mind, however, and that
arises from the fact that when one has an incident lepton, a
scattering with exchange of a $\gamma$, $W^\pm$ or $Z^0$, and
detects the scattered lepton (i.e., a charged lepton), the
$t$-channel exchange of the appropriate boson is controlled. In
contrast, when the scattered lepton is a neutrino or antineutrino,
and therefore not detected, but instead a knocked-out nucleon is
detected, it is the $u$-channel whose kinematics are controlled
(see also~\cite{Barbaro:1996vd} for discussions of this case).
Accordingly, in the scaling analysis it is not obvious that the
two types of processes are simply related, and therefore to apply
the scaling ideas to NC neutrino and antineutrino scattering, in
particular for discussions of differential cross sections as in
the present work, we first have to address the issue of how the
$t$- and $u$-channels are related.

The paper is organized the following way: in Sec.~\ref{sec:sec2}
we begin with a basic discussion of the $t$- and $u$-channel
kinematics involved in the semi-leptonic electroweak processes of
interest (Subsec.~\ref{subsec:subsec2a}) followed by a brief
summary in Subsec.~\ref{subsec:subsec2b} of the cross section
formalism and the ideas of scaling when inter-relating $t$- and
$u$-channel processes. To keep the discussions relatively brief in
this subsection, the development of the single-nucleon NC neutrino
and antineutrino cross sections is placed in an Appendix. For
orientation in Subsec.~\ref{subsec:subsec2c} the Relativistic
Fermi Gas (RFG) model is invoked and its superscaling properties
summarized. Then in Sec.~\ref{sec:sec3} our results are presented
and our conclusions are gathered in Sec.~\ref{sec:sec4}.

\section{General formalism for u-channel scattering}
\label{sec:sec2}

We begin the general discussion of how $t$- and $u$-channel
semi-leptonic reactions are inter-related with a summary of the
relevant kinematic variables in the problem.

\subsection{Kinematics}
\label{subsec:subsec2a}

\begin{figure}[th]
\begin{center}
\includegraphics[scale=0.9, bb=140 460 400 710]{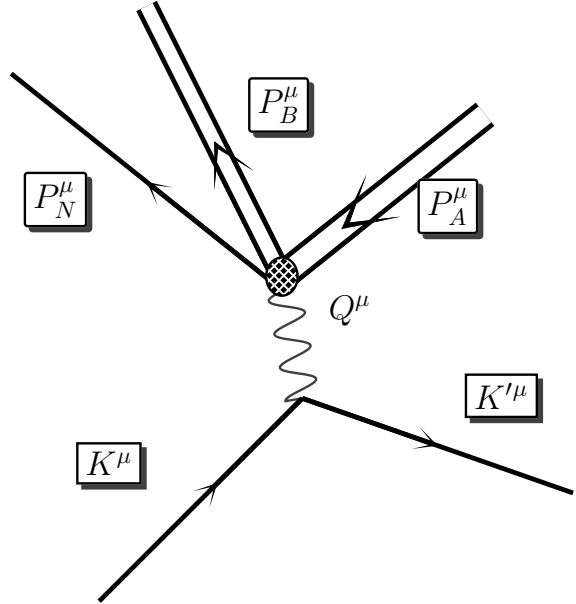}
\caption{Kinematics for semi-leptonic nucleon
knockout reactions in the one-boson-exchange approximation.}
\label{fig0}
\end{center}
\end{figure}

We consider general semi-leptonic quasi-free scattering from
nuclei in Born approximation.

We start with one basic assumption that is usually presumed to be
a good approximation in the kinematic region where quasielastic
scattering is dominant, namely, that the inclusive cross sections
are well represented by the sum of the integrated semi-inclusive
proton and neutron emission cross sections. In doing so we are
neglecting processes that occur for the same kinematics, but have
{\em no emitted nucleon in the final state} (photon emission,
deuteron emission, alpha emission, coherent pion production, etc.,
but without an emitted nucleon). The process of interest (see
Fig.~\ref{fig0}) has a lepton of 4-momentum $K^{\mu }=(\epsilon
,\mathbf{k})$ scattered to another lepton of 4-momentum $K^{\prime
\mu }=(\epsilon ^{\prime },\mathbf{k}^{\prime })$, exchanging a
vector boson with 4-momentum $Q^{\mu }=K^{\mu }-K^{\prime \mu }$.
The lepton energies are $\epsilon =\sqrt{m^{2}+k^{2}}$ and
$\epsilon ^{\prime } =\sqrt{m^{\prime 2}+k^{\prime 2}}$, with $m$
$(m^{\prime })$ the mass of the initial (final) lepton. For
NC neutrino scattering $m=m^{\prime }=0$
(assuming zero-mass neutrinos). Note that no assumption such as
the plane-wave impulse approximation is being invoked at this
stage.

In the laboratory system the initial nucleus is in its ground
state with 4-momentum $P_{A}^{\mu } =(M_{A}^{0},0)$. The final
hadronic state corresponds to a nucleon ($N=p$ or $n$) with
4-momentum $P_{N}^{\mu }=(E_{N},\np_N)$ and energy
$E_{N}=\sqrt{m_{N}^{2}+p_{N}^{2}}$ plus an unobserved daughter
nucleus with 4-momentum $P_{B}^{\mu }=(E_{B},\mathbf{p}_{B})$. As
usual in semi-leptonic reactions we introduce the missing momentum
$\mathbf{p} \equiv -\mathbf{p}_{B}$ and the excitation energy
$\mathcal{E}\equiv  E_{B} - E_{B}^{0}$, with $E_{B}^{0}
=\sqrt{\left( M_{B}^{0}\right) ^{2}+p^{2}}$, $M_{B}^{0}$ being the
ground-state mass of the daughter system (for details
see~\cite{Day:1990mf,Donnelly:1998xg,Donnelly:1999sw}).

\begin{figure}[th]
\begin{center}
\includegraphics[scale=0.95,  bb= 140 520 470 710]{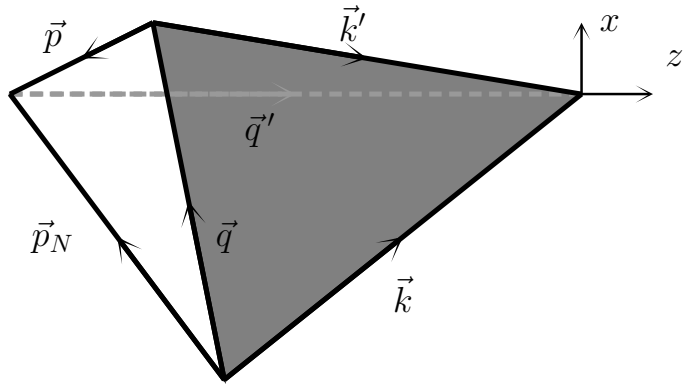}
\caption{Vectors relating $t$-channel and
$u$-channel kinematic variables.} \label{fig0c}
\end{center}
\end{figure}

For NC neutrino scattering we assume that the neutrino beam
momentum is specified and the outgoing nucleon is detected. Hence
$p_{N}$ and the angle $\theta _{kp_{N}}$ (between $\mathbf{k}$ and
$\mathbf{p}_{N}$) are given. Note that the scattered lepton's
4-momentum is not specified, as would be the case for $t$-channel
scattering. In analogy with the $t$-channel case, we can define a
$u$-channel exchanged 4-momentum
\begin{eqnarray}
Q^{\prime \mu } &\equiv &K^{\mu }-P_{N}^{\mu }
=(\omega ^{\prime },\mathbf{q}^{\prime }).  \label{e9}
\end{eqnarray}
The above equation yields
\begin{equation}
q^{\prime }=|\mathbf{q}^{\prime }|=\sqrt{k^{2}+p_{N}^{2}-2kp_{N}\cos \theta
_{kp_{N}}}\ .  \label{e15}
\end{equation}
For convenience in looking at the kinematics one can use a coordinate system
having the z-axis along $\mathbf{q}^{\prime }$, with $%
\mathbf{k}$ and $\mathbf{p}_{N}$ lying in the xz-plane. The
vectors $\mathbf{k}^{\prime }$ and $\mathbf{p}=\mathbf{k}^{\prime
}-\mathbf{q}^{\prime }$ lie in a plane forming an angle
$\phi^{\prime }$ with the xz-plane defined above (see
Fig.~\ref{fig0c}).

The exclusive process illustrated in Fig.~\ref{fig0} is fully
determined by six kinematic variables, which can be chosen to be
($k$, $p_{N}$, $\theta _{kp_{N}}$, $p$, $\mathcal{E}$,
$\phi^{\prime }$). The $u$-channel inclusive cross section for
$(k,p_{N},\theta _{kp_{N}})$ fixed is obtained by integrating over
the allowed region in the ($p$, ${\mathcal{E}}$)-plane and over
the azimuthal angle, $0\leq \phi ^{\prime }\leq 2\pi $. Again
referring to Fig.~\ref{fig0c}, one sees that at fixed $u$-channel
scattering kinematics (i.e., the triangular region bounded by
$\mathbf{k}$, $\mathbf{p}_N$ and $\mathbf{q'}$ fixed) and for a
given point in the ($p$, ${\mathcal{E}}$)-plane, this $\phi
^{\prime }$ integration corresponds to having the triangle bounded
by $\mathbf{p}$, $\mathbf{k'}$ and $\mathbf{q'}$ fixed in size and
shape but rotating about the $z$-axis, namely, through the full
range of the azimuthal angle $\phi ^{\prime }$. This clearly
implies that the $t$-channel momentum transfer $q$ varies and that
the usual azimuthal angle $\phi$ (rotations about $\mathbf{q}$)
does not cover the full range $(0,2\pi)$. This has consequences
that are discussed in more detail below.

In order to determine the integration region in the ($p$,
${\mathcal{E}}$)-plane we use energy conservation, obtaining the
following expression:
\begin{equation}
{{\mathcal{E}}}=(M_{A}^{0}+\omega ^{\prime })-\left[ \sqrt{m^{\prime
2}+q^{\prime 2}+p^{2}+2q^{\prime }p\cos \theta _{q^{\prime }p}}+\sqrt{\left(
M_{B}^{0}\right) ^{2}+p^{2}}\right] .  \label{e25}
\end{equation}
Following the usual y-scaling analysis we can now examine the various curves
${\mathcal{E}}={\mathcal{E}}(p)$ that result when various choices are made
for $%
\cos \theta _{q^{\prime }p}$. Let us call the curves ${\mathcal{E}}_{\pm
}^{\prime }(p)$ when $\cos \theta _{q^{\prime }p}=\pm 1:$%
\begin{equation}
{\mathcal{E}}_{\pm }^{\prime }(p)=(M_{A}^{0}+\omega ^{\prime })-\left[ \sqrt{%
m^{\prime 2}+(q^{\prime }\pm p)^{2}}+\sqrt{\left( M_{B}^{0}\right) ^{2}+p^{2}%
}\right] .  \label{e26}
\end{equation}

  From these we can proceed to
find the intersections of the curves with the axis ${\mathcal{E}}=0$. This
leads to definitions for a scaling variable $y^{\prime }$ and a
maximum missing momentum $Y^{\prime }$:
\begin{eqnarray}
y^{\prime } &\equiv &\frac{1}{W_{X}^{2}}\left[ (M_{A}^{0}+\omega ^{\prime })%
\sqrt{\Lambda _{X}^{2}-\left( M_{B}^{0}\right) ^{2}W_{X}^{2}}-q^{\prime
}\Lambda _{X}\right]  \label{e27} \\
Y^{\prime } &\equiv &\frac{1}{W_{X}^{2}}\left[ (M_{A}^{0}+\omega ^{\prime })%
\sqrt{\Lambda _{X}^{2}-\left( M_{B}^{0}\right) ^{2}W_{X}^{2}}+q^{\prime
}\Lambda _{X}\right] ,  \label{e28}
\end{eqnarray}
where
\begin{eqnarray}
W_{X} &=&\sqrt{(M_{A}^{0}+\omega ^{\prime })^{2}-q^{\prime 2}}  \label{e29}
\\
\Lambda _{X} &=&\frac{1}{2}\left( W_{X}^{2}+\left( M_{B}^{0}\right)
^{2}-m^{\prime 2}\right) .  \label{e30}
\end{eqnarray}
Note that these are new variables and not simply related to the
variables $y$ and $Y$ that come from the familiar y-scaling
analysis~\cite{Day:1990mf,Barbaro:1998gu,Donnelly:1998xg,Donnelly:1999sw}.
The allowed region is then determined: for $y^{\prime }<0$
one has $-y^{\prime }\leq p\leq Y^{\prime }$ with $0\leq {\mathcal{E}}\leq
{\mathcal{E}}_{-}^{\prime }(p)$, whereas for $y^{\prime }>0$ one has for $%
0\leq p\leq y^{\prime }$ the range ${\mathcal{E}}_{+}^{\prime
}(p)\leq {\mathcal{E}}\leq {\mathcal{E}}_{-}^{\prime }(p)$, and
for $y^{\prime }\leq p\leq Y^{\prime }$ the range $0\leq
{\mathcal{E}}\leq {\mathcal{E}}_{-}^{\prime }(p)$. When
$y^{\prime}=0$ one covers the largest range in missing momentum at
the minimal missing energy and accordingly somewhere near this
point the inclusive integral is expected to be at a maximum;
namely, this kinematic point corresponds approximately to the QE
peak.

Concerning the azimuthal integration, note that kinematic
variables entering the usual $t$-channel (such as the momentum and
energy transfer $(q,\omega)$, the lepton scattering angle $\theta
_l$ between $\nk$ and $\nk'$ and the solid angle defining the
outgoing nucleon momentum $(\theta _{qp_{N}},\phi_N)$) all depend
on $\cos \phi ^{\prime }$ --- see the above discussions of
Fig.~\ref{fig0c}. Thus the integration over $\phi ^{\prime }$
implies an integration over the azimuthal angle $\phi _{N}$;
however, as $\phi ^{\prime }$ varies, the integration implied over
$\phi _{N}$ is not being done at constant $(q,\omega) $.
Furthermore, the range over which the implied $\phi
_{N}$-integration occurs is not in general the full range. This
implies that the symmetry properties of the responses $R_K$
cannot be used in the case of $u$-channel inclusive scattering to
eliminate some of the responses (e.g., the TL and TT terms), as is
the case for $t$-channel inclusive scattering.

\subsection{Cross sections and scaling}
\label{subsec:subsec2b}

Next we turn to a discussion of the basic cross sections and
scaling variables involved in the present study. As discussed
above we consider only semi-inclusive nucleon knockout reactions
in building up the inclusive cross sections. The usual
procedure~\cite{Barbaro:1996vd} is to start with the Plane Wave
Impulse Approximation (PWIA) for the $(l,l' N)$ cross section and
integrate over all unconstrained kinematic variables. Final-state
interactions are then presumed to occur after the primary
electroweak interaction with a nucleon in the nucleus and so, for
instance, a succession of $(N,2N)$ steps occurring during the time
evolution of the high-energy emitted nucleon as it proceeds
through the nuclear medium can cause a redistribution of strength
in the missing-energy, missing-momentum plane (see~\cite{Nieves05}
for recent work along these lines).  Such processes tend to move
strength from lower missing energies to higher ones and thereby
produce an asymmetry in the scaling function, skewing it to larger
values of energy loss $\omega$ or, equivalently, in the positive
scaling variable direction.  Other
approaches~\cite{Caballero:2005sj} also yield an asymmetric
scaling function --- in agreement with experiment
--- when strong final-state interactions are incorporated, again
via a shift of strength to higher missing energies.

In contrast, in the present work where our emphasis is placed on
inter-relating various inclusive semi-leptonic processes, and not
on detailed modeling of the reaction chain, we take as given the
full semi-inclusive nucleon knockout cross section (i.e., given by
Nature) and proceed to integrate to obtain inclusive cross
sections. Clearly this does not imply that we have a full
understanding of the former, only that asymptotic states may be
used to account for all open channels and that it is not necessary
to account for the entire sequence of steps that yields these
states.

For $t$-inclusive scattering, where $Q^\mu\equiv(\omega,\nq)$ is
constant and the final lepton is detected (as in usual inclusive
electron scattering or in charge-changing neutrino reactions), the
inclusive cross section is calculated by integrating the
semi-inclusive cross section $d\sigma / d\Omega_{k'} dk' d\Omega_N
dp_N$ over the ejected nucleon (and summing over protons and
neutrons), whereas for the $u$-inclusive scattering we are
considering here, where $Q'^\mu\equiv(\omega',\nq')$ is constant
and the final nucleon is detected, one has to integrate over the
final lepton. That is we have
\ba \frac{d\sigma}{d\Omega_{k'} dk'
} &=& \int d\Omega_N dp_N \frac{d\sigma}{d\Omega_{k'} dk' d\Omega_N dp_N} \label{newt} \\
\frac{d\sigma}{d\Omega_N dp_N}&=& \int d\Omega_{k'} dk'
\frac{d\sigma}{d\Omega_{k'} dk' d\Omega_N dp_N} \label{newu} \ea
for $t$- and $u$-channel reactions, respectively. These integrals
can be transformed into integrals in the $(p,{\cal E})$-plane
using the relations \ba d\Omega_{N} dp_N &=&
\left(\frac{E_N}{p_N^2}\right)\frac{1}{q} p dp d{\cal E} d\phi_N
\\
d\Omega_{k'} dk' &=&
\left(\frac{\epsilon'}{k'^2}\right)\frac{1}{q'} p dp d{\cal E}
d\phi'\ . \ea This leads to the following expressions for the
inclusive cross sections, in the $t$-channel \ba
\frac{d\sigma}{d\Omega_{k'} dk'} &=& \frac{2\pi}{q} \int_{{\cal
D}_t} p dp \int d{\cal E} \int_0^{2\pi} \frac{d\phi_N}{2\pi}
\left(\frac{E_N}{p_N^2}\right) \frac{d\sigma}{d\Omega_{k'} dk'
d\Omega_N dp_N}
 \label{eq:inclt} \ea and
in the $u$-channel \ba \frac{d\sigma}{d\Omega_{N} dp_N} &=&
\frac{2\pi}{q^\prime} \int_{{\cal D}_u} p dp \int d{\cal E}
\int_0^{2\pi} \frac{d\phi'}{2\pi}
\left(\frac{\epsilon'}{k'^2}\right) \frac{d\sigma}{d\Omega_{k'}
dk' d\Omega_N dp_N} \ , \label{eq:inclu} \ea respectively. The
above expressions are simply connected to one other by
interchanging the final lepton variables with the final nucleon
variables, but for the fact that the integration regions
 ${\cal D}_t$ and ${\cal D}_u$
in the $(p,{\cal E})$-plane are different in the two cases.
The $t$-channel case is discussed in~\cite{Barbaro:1996vd}, while
the $u$-channel case is treated in the following section on
results.

To this point we have made only relatively weak approximations by
assuming that the cross sections in the quasielastic region are
dominated by integrals over the semi-inclusive nucleon knockout
cross sections. Following~\cite{Barbaro:1996vd} we write the
latter in terms of products of single-nucleon electroweak cross
sections multiplied by what may be called the {\em reduced cross
section} $\Sigma$: \be \frac{d\sigma}{d\Omega_{k'} dk' d\Omega_N
dp_N} = \frac{1}{(2\pi)^2}\frac{1}{2\epsilon}\frac{1}{2E} g^4
D_V(Q^2)^2 l_{\mu\nu}w^{\mu\nu}
\left(\frac{k'^2}{2\epsilon'}\right)
\left(\frac{p_N^2}{2E_N}\right)
\Sigma(q,\omega,\theta_{kk'},\phi',p,{\cal E})\ , \label{eq:semi1}
\ee where $E$ is the energy of the struck nucleon, $g$ is the
strength of the fermion-vector boson coupling and
$D_V(Q^2)=(Q^2-M_V^2)^{-1}$ is the vector boson propagator,
 while $l_{\mu\nu}$ and $w^{\mu\nu}$ are the
usual leptonic and (single-nucleon) hadronic tensors, respectively.
Clearly other sets of independent variables may be used as
arguments of the reduced cross section (see below).

Next we make two stronger approximations. First, we assume that
the single-nucleon cross section varies only slowly with $(p,{\cal
E})$ and may be removed from the integrals over $p$ and ${\cal
E}$. This has been verified for $t$-channel reactions as long as
the semi-inclusive cross sections are peaked at low missing-energy
and missing-momentum (see, for example, \cite{Day:1990mf}). In
particular, for the $t$-inclusive case the vector boson propagator
can be extracted from the integral, and the same applies to the
single-nucleon form factors appearing in $w^{\mu\nu}$, as they
only depend upon $Q^2$. As a consequence, in $t$-channel case one
can verify that the $(p,{\cal E})$ dependence of the
single-nucleon cross section is weak at constant $(\omega,q)$ and
therefore its mean value (namely, integrated over $\phi_N$ and
divided by $2\pi$) can be removed from the integrations in
Eq.~(\ref{eq:inclt}). The $u$-channel case is more complicated and
is dealt with below.

If we make this approximation we are left with \be
\frac{d\sigma}{d\Omega_{k'} dk'} \simeq
\overline{\sigma}_{sn}^{(t)} F(y,q) \ , \ee where \be F(y,q)
\equiv \int_{{\cal D}_t} p dp \int \frac{d{\cal E}}{E}
\Sigma(q,\omega,\theta_{kk'},\phi',p,{\cal E}) \label{eq:scf} \ee
depends upon the scaling variable $y$ and the momentum transfer
$q$~\cite{Donnelly:1999sw,Donnelly:1998xg,Day:1990mf,Barbaro:1998gu,Alberico:1988bv}.
Note that the reduced cross section $\Sigma$ occurring above would
be the spectral function $S(p,{\cal E})$, namely dependent only on
$(p,{\cal E})$, were the PWIA to be assumed; however, no such
assumption is being made here.

Second, we assume {\em factorization} in the sense that the
reduced cross section appearing above depends only weakly on the
momentum transfer $q$, this dependence being contained mostly in
the single-nucleon cross section. Note that, for instance,
residual dependence in $\Sigma$ on the scaling variable $y$ is not
part of the factorization assumption. Such dependence would not
lead to any scaling violation. This means that factorization is
not equivalent to assuming dependence only on $(p,{\cal E})$ as in
the PWIA. Clearly missing here, for instance, are processes
involving meson-exchange currents~\cite{Amaro02}--\cite{DePace04}
which in this sense do not factor, as their dependences on $q$ are
clearly not the same as those contained in the single-nucleon
cross section which has been divided out to define the reduced
cross section. However, our past studies of superscaling show
that, for high-energy inclusive scattering at quasielastic
kinematics, the scaling behaviour is quite well respected, with
perhaps 10\% or so left to be explained by effects such as those
from MEC which should break the scaling. Indeed, even with
relatively strong final-state interactions one finds in some
modeling~\cite{Caballero:2005sj} that the scaling is maintained,
suggesting that the above assumption is valid, at least in the
region of the QE peak.

We note in passing that the violations of scaling of the first kind, 
namely, some residual dependence on the momentum transfer $q$, even 
at the level of the above equation can stem from two different sources:
(1) the region of integration ${\cal D}_t$ depends on $q$ and only for
asymptotically high $q$ does it approach a $q$-independent form, and (2)
the reduced cross section $\Sigma$ may contain some weak dependence on
$q$. Indeed, from the observation that the approach to first-kind scaling
is from above, i.e. the measured reduced cross section decreases with $q$ 
before reaching the scaling domain (see~\cite{Day:1990mf}, for example), 
it appears that (2) must occur, and not just (1) which would imply an 
approach from below, since the integration region increases with $q$.

At high energies, where the scaling idea works and scaling of
first kind is reasonably good, we find that $F(y,q)\simeq
F(y)\equiv F(y,\infty)$ and is not a function of $q$, in effect
validating the factorization assumption and the quality of the
approximation where a mean-value for the single-nucleon cross
section is removed from the integrals. This was used
in~\cite{Amaro:2004bs} to predict the charge-changing (CC)
neutrino cross section: we let Nature solve for us the integral in
Eq.~(\ref{eq:scf}) to obtain an empirical function $F(y)$ from
electron scattering to be used in CC neutrino studies.

In the $u$-inclusive case the above factorization is not trivial,
since $Q^2$ varies within the integration region. However, one can
again assume that \be \label{crossu} \frac{d\sigma}{d\Omega_{N}
dp_N} \simeq \overline{\sigma}_{sn}^{(u)} F(y',q') \ , \ee where
\be F(y',q') \equiv \int_{{\cal D}_u} p dp \int \frac{d{\cal
E}}{E} \Sigma \simeq F(y')~, \label{eq:scfu} \ee provided the
effective NC single-nucleon cross section \be
\overline{\sigma}_{sn}^{(u)}= \frac{1}{32\pi
\epsilon}\frac{1}{q^\prime} \left(\frac{p_N^2}{E_N}\right) g^4
\int_0^{2\pi} \frac{d\phi'}{2\pi}
l_{\mu\nu}(\nk,\nk')w^{\mu\nu}(\np,\np_N) D_V(Q^2)^2
\label{crossu1} \ee is almost independent of $(p,{\cal E})$ for
constant $(k,p_N,\theta_{kp_N})$. This seems indeed to be the
case, as shown from numerical studies presented below in the
Results section. Then, as in~\cite{Amaro:2004bs}, the empirically
determined scaling function $F(y')$ can be used to predict
realistic NC cross sections.

To be able to use the scaling function obtained from analyses of inclusive
electron scattering data for predictions of neutrino reaction cross 
sections one further assumption must be made, namely, the domains of
integration in the integrals above must be the same or at least very similar. 
In the case of CC neutrino reactions this is clearly the case except
at very low energies for the muon case where the kinematic dependence
on the muon mass is important in determining ${\cal D}_t$. For NC neutrino
reactions the integration domain ${\cal D}_u$ differs to some degree from
the one that enters in electron scattering, namely, ${\cal D}_t$. In particular,
when determining the scaling function $F(y',q')$ with input from electron 
scattering which yields $F(y,q)$, clearly the first step is to use the latter
evaluated at $y=y'$ and to work in the scaling regime where $q$ and $q'$ are
both large enough to make the regions in the $(p,{\cal E})$-plane extend to
high-$p$ and high-${\cal E}$ (see the arguments for electron scattering scaling
summarized, for instance, in~\cite{Day:1990mf}). Under these circumstances the regions
denoted ${\cal D}_t$ and ${\cal D}_u$ differ significantly only at large ${\cal E}$
(also at large $p$, but there one believes the semi-inclusive cross sections are
negligible). Accordingly, given that the semi-inclusive cross sections are dominated
by their behaviours at low ${\cal E}$ and low $p$, one expects the results of the integrations
in the two cases, $t$-channel and $u$-channel, to be very similar, and thus
the scaling functions will be essentially the same. Were this not to be the case, then
it would be likely that first-kind scaling for inclusive electron scattering would not 
occur, in contradiction with observation. 

A further difference between the $t$- and $u$-scattering cases
should be stressed. In both cases the single-nucleon cross section
can be expressed in terms of response functions, as shown in the
Appendix. However, as mentioned above, for $t$-inclusive processes
the special symmetry about the $\nq$ direction can be exploited to
remove the TL, TT and TL$^{\prime}$ responses after performing the
$\phi_N$-integration, which simply yields a factor $2\pi$. In the
$u$-channel, instead, the unrestricted integration over $\phi'$
yields an effective integration over $\phi_N$ which is not uniform
and does not in general cover all of the interval $(0,2\pi)$. As a
consequence the TL, TT and TL$^{\prime}$ responses do contribute.
As we will show later, their contribution is suppressed and only
the TL contribution is relevant for the kinematics of interest in
the present study.

\subsection{RFG and superscaling}
\label{subsec:subsec2c}

In this section we discuss the NC neutrino cross section in the
RFG model, which corresponds to the following excitation energy
\be {\cal E}_{RFG}(p)=\sqrt{m_N^2+k_F^2}-\sqrt{m_N^2+p^2}
\label{efermi} \ee and spectral function  \be S_{RFG}(p,{\cal E})=
\frac{3k_F}{4T_F}\theta(k_F-p)\delta({\cal E}-{\cal E}_{RFG}(p)) \
, \label{eq:SRFG} \ee where $k_F$ is the Fermi momentum and
$T_F=\sqrt{k_F^2+m_N^2}-m_N$ the Fermi kinetic energy.

Due to the delta-function in Eq.~(\ref{eq:SRFG}) the integration
region in the $(p,{\cal E})$-plane simply reduces to a line and
the lower limit in the integral over $p$ is given by the intercept
of the curve ${\cal E}_{RFG}(p)$ with ${\cal E}'_{-}(p)$ when
$y'<0$. When $y'>0$ it is given by the intercept of ${\cal
E}_{RFG}(p)$ with ${\cal E}'_{+}(p)$ (${\cal E}'_{-}(p)$) when
${\cal E}'_{\pm}(0)<T_F$ (${\cal E}'_{\pm}(0)>T_F$). By solving
these equations it is easily shown that the minimum momentum
required for a nucleon to participate in the reaction is \be
p_{\rm min} = \left|y^{(u)}_{RFG}\right| \ , \ee where \be
y^{(u)}_{RFG}= s \frac{m_N}{\tau'} \left[
\lambda'\sqrt{\tau'^2\rho'^2+\tau'}-\kappa'\tau'\rho'\right] \ee
is the RFG y-scaling variable associated with $u$-scattering
(hence the index $(u)$ to distinguish it from the usual
$t$-channel variable). Moreover we have introduced the
dimensionless kinematic quantities $\kappa'\equiv q'/2 m_N$,
$\lambda'\equiv \omega'/2 m_N$, $\tau'=\kappa'^2-\lambda'^2$ and
defined $\rho'\equiv 1-\frac{1}{4\tau'} (1-m'^2/m_N^2)$. The sign
$s$ is \be s\equiv{\rm sgn}\left\{ \frac{1}{\tau'} \left[
\lambda'\sqrt{\tau'^2\rho'^2+\tau'}-\kappa'\tau'\rho'\right]\right\}
\ . \ee As in electron scattering, it is convenient to introduce a
dimensionless scaling variable
\be
\psi^{(u)}_{RFG}=s\sqrt{\frac{m_N}{T_F}}
\left[\sqrt{1+\left(\frac{y^{(u)}_{RFG}}{m_N}\right)^2}-1\right]^{1/2}~,
\ee representing the minimum kinetic energy of the nucleons
participating in the reaction. By placing the spectral function of
Eq.~(\ref{eq:SRFG}) in Eq.~(\ref{eq:scfu}) one immediately finds
the RFG scaling function \be F_{RFG}(\psi^{(u)}_{RFG}) = \frac{3
k_F}{T_F}\int_{E_{min}}^{E_F} dE \int d{\cal E} \delta\left({\cal
E}-{\cal E}_{RFG}\right) =\frac{3}{4}
k_F\left(1-\psi^{(u)2}_{RFG}\right)
\theta\left(1-\psi^{(u)2}_{RFG}\right)~. \label{eq:scfuRFG} \ee

Providing the single-nucleon cross section is smoothly varying
within the $(p,{\cal E})$ integration region, the differential RFG
cross section can be factorized as shown in Eq.~(\ref{crossu})
with the scaling function given by Eq.~(\ref{eq:scfuRFG}). More
realistic predictions can be given by using, instead of the RFG
scaling function, the empirical scaling function as determined
from QE electron scattering, as already done
in~\cite{Amaro:2004bs} for charged current reactions. These are
discussed in the next section.

\section{Results}
\label{sec:sec3}

Before presenting our predictions for the cross section, we test
the validity of the scaling approach in the $u$-channel. To this
end we analyze how the effective NC single-nucleon cross section
$\overline{\sigma}_{sn}^{(u)}$ given in Eq.~(\ref{crossu1})
depends on the missing momentum $p$ and excitation energy ${\cal
E}$ for selected values of the kinematical variables
$(k,E_N,\theta_{kp_N})$. To proceed, we assume the proton knockout
case and divide $\overline{\sigma}_{sn}^{(u)}$ evaluated in the
whole $(p,{\cal E})$-plane by its value corresponding to $p=|y'|$
and ${\cal E}=0$. In what follows we use 
the cc2 off-shell prescription for the nucleon current and 
the H\"{o}hler parameterization for the 
single-nucleon form factors~\cite{Hoehler:1976}, ignoring the strangeness 
content of the nucleon, unless specified otherwise.

\begin{figure}[th]
\begin{center}
\includegraphics[scale=0.95, bb=70 540 500 780]{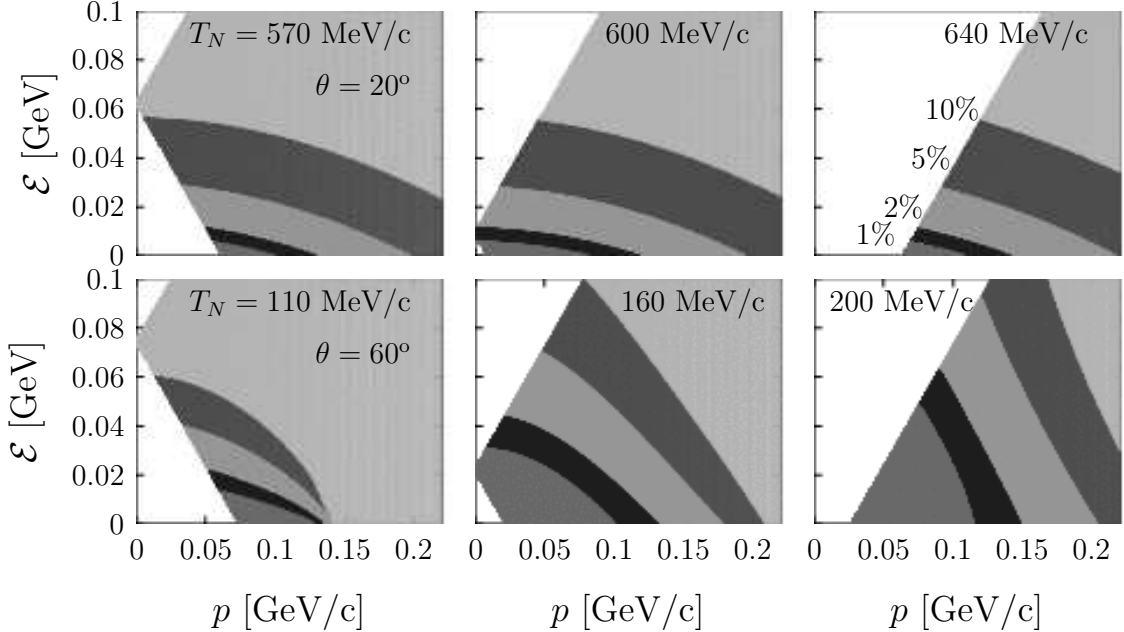}
\caption{ The various regions correspond to values of the ratio
${\cal R}\equiv \overline{\sigma}_{sn}^{(u)}(p,{\cal
E})/\overline{\sigma}_{sn}^{(u)}(p=|y'|,{\cal E}=0)$ differing
from 1 by at most $1\%$ (lowest region), $1-2\%$, $2-5\%$,
$5-10\%$ and more than $10\%$ (highest region). For this figure
proton knockout has been assumed; the neutron knockout case is
similar and not shown. 
For brevity, in this figure we let $\theta$ 
stand for the angle $\theta_{kp_N}$. 
\label{fig1}}
\end{center}
\end{figure}

The results are given in Fig.~\ref{fig1} in terms of different
shadings representing the regions where this ratio differs from
unity by at most 1\%, 1--2\%, 2--5\%, 5--10\% and more than 10\%
respectively, as indicated in the top right panel. The six graphs
correspond to two values of the scattering angle $\theta_{kp_N}$:
$20^0$ (top panels) and $60^0$ (bottom panels). In each case, the
outgoing proton kinetic energies have been selected to correspond
to the regions below, above and close to the peak of the
differential cross section. While not shown here, the results for
neutron knockout are very similar to those for proton knockout.

The results in Fig.~\ref{fig1} illustrate the validity of the
scaling approach. Only for very large values of the
excitation energy does the effective NC single-nucleon cross
section depend significantly on $(p,{\cal E})$. In fact,
restricting ourselves to excitation energies below twice the
maximum value of the RFG model, ${\cal E}\simeq 50$ MeV, the
dispersion presented by the ratio is at most $\sim$5--10\%.

\begin{figure}[th]
\begin{center}
\includegraphics[scale=0.75, bb= 50 400 540 730]{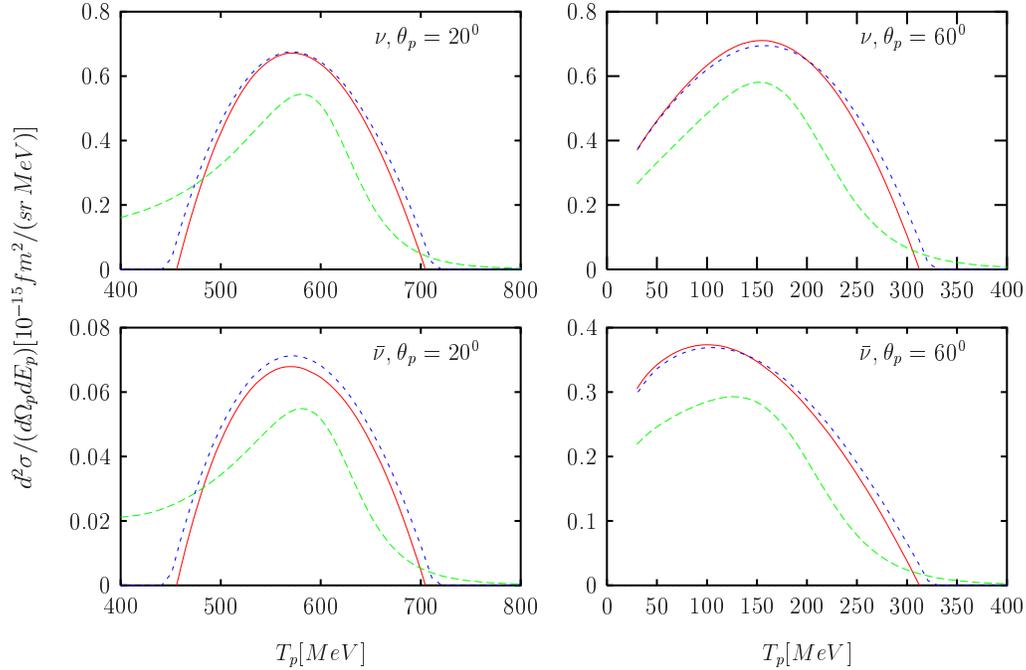}
\caption{(Color online) Quasielastic differential cross section for neutral
current neutrino and antineutrino scattering at 1 GeV from
$^{12}$C for proton knockout obtained using the RFG
(short-dashed), the factorized approach with the RFG scaling
function (solid) and the empirical scaling function (long-dashed).
For brevity, in this and in the following figures we let $\theta_N$ 
stand for the angle $\theta_{kp_N}$. 
\label{fig2p}}
\end{center}
\end{figure}

\begin{figure}[th]
\begin{center}
\includegraphics[scale=0.75, bb= 50 400 540 730]{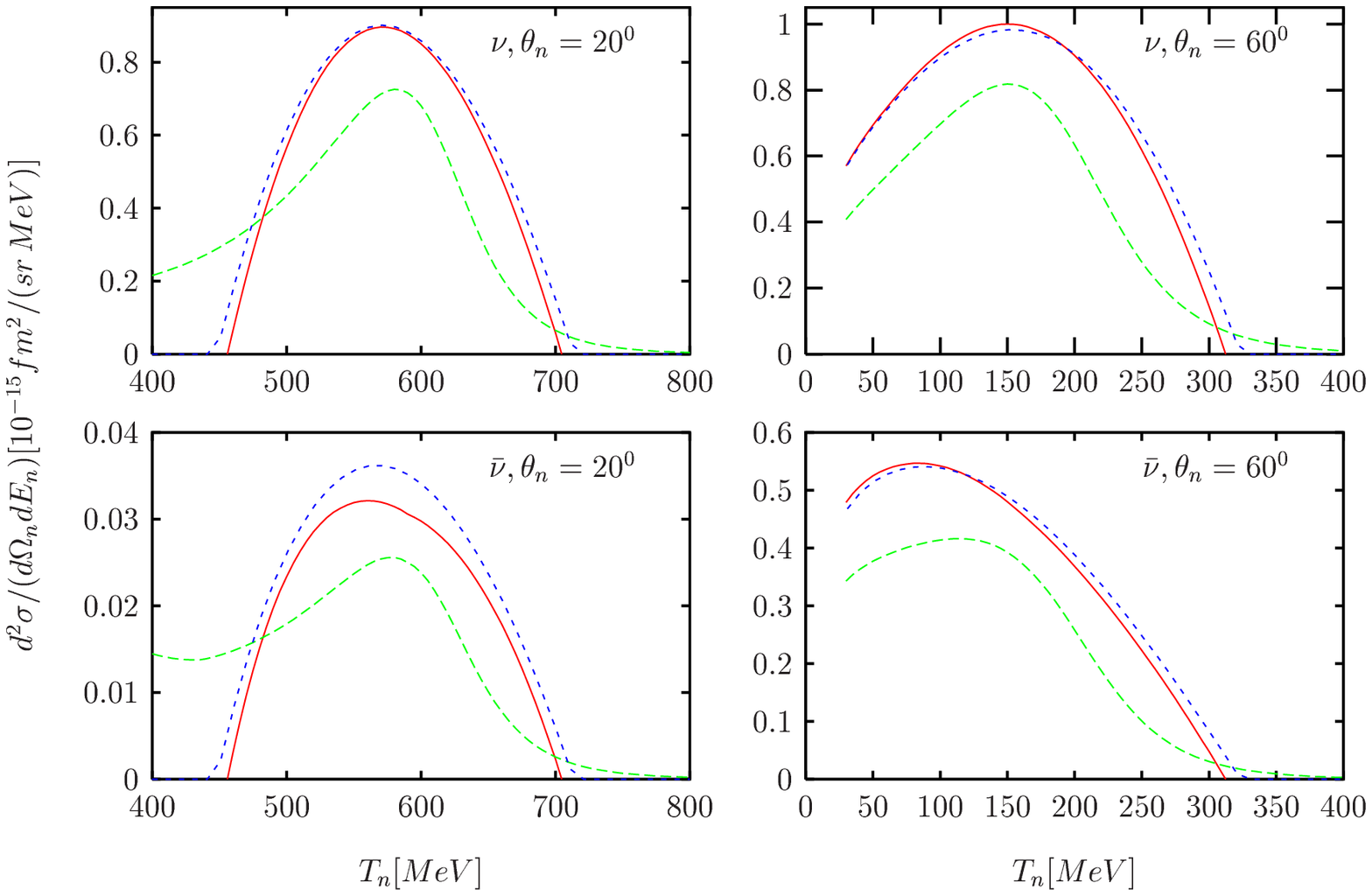}
\caption{(Color online) As for the previous figure, but now
showing the neutron knockout case. \label{fig2n}}
\end{center}
\end{figure}

This outcome is also in accordance with the results presented in
Figs.~\ref{fig2p} (proton case) and \ref{fig2n} (neutron case).
Here we show the neutral current neutrino (upper panels) and
antineutrino (lower panels) double differential cross sections for
scattering at 1 GeV from $^{12}$C as a function of the ejected
proton or neutron kinetic energy. The scattering angles have been
fixed as in the previous figure.

Beginning with the RFG model, as in past work the Fermi momentum
for $^{12}$C is taken to be $k_F=228$ MeV/c and results are given
both using the full RFG model (short-dashed curves) and making use
of the factorization approach assumed in Eq.~(\ref{crossu}) with
the $u$-channel NC single-nucleon cross section evaluated at
$p=y_{RFG}$ and ${\cal E}={\cal E}_{RFG}$ (solid lines). One sees
that the two sets of results almost coincide in the whole $T_N$
region where the RFG cross section is defined, indicating that the
scaling argument works very well.

Hence we may use the phenomenological scaling function extracted
from $(e,e')$ data, as was done in our previous CC neutrino
reaction analysis~\cite{Amaro:2004bs}, to predict NC
neutrino-nucleus scattering cross sections. These are also plotted
in Figs.~\ref{fig2p} and \ref{fig2n} as long-dashed lines: they
are seen to be lower by about $25\%$ at the peak than the RFG
results, an effect similar to what was found
in~\cite{Amaro:2004bs} for the charge-changing processes.
Moreover, the empirical scaling function leads to cross sections
extending both below and above the kinematical region where the
RFG is defined. In particular, the long tail displayed for low
$T_N$-values (corresponding to positive values of the scaling
variable $\psi'$) is noteworthy. This tail arises not only from
the asymmetric shape of the phenomenological scaling function, but
also from the effective NC single-nucleon cross section, which
increases significantly for low $T_N$ values.

On comparing Figs.~\ref{fig2p} and \ref{fig2n} we see that the
shapes of the cross sections for proton and neutron knockout are
very similar, although the magnitudes are somewhat different:
except for antineutrinos at forward angles, where the cross
sections are very small, the neutron knockout results are 30--50\%
higher than for proton knockout. This occurs because (in absence
of strangeness) both the vector and the axial-vector contributions
are larger for neutrons than for protons, and they sum up. In
particular, the AA piece is the same for $p$ and $n$, since
$\widetilde G_{Ap}= -\widetilde G_{An}$ (see Eq.~(A33)). On the
other hand, from Eqs.~(A28,A29,A34,A35) one has that $\widetilde
G_{Ep} \simeq -G_{En}$ and $\widetilde G_{En} \simeq -G_{Ep}$.
Hence, when compared with electromagnetic interactions, the roles
of protons and neutrons are reversed in the weak neutral sector,
so that $|\widetilde G_{En}|>>|\widetilde G_{Ep}|$. Similarly,
from Eqs.~(A30,A31,A36,A37) one finds that $|\widetilde
G_{Mn}|>|\widetilde G_{Mp}|$. For antineutrinos things are more
delicate, since the VA response has the opposite sign. For
instance, for neutron knockout at $\theta_n=20^0$ the sum VV+AA
almost exactly cancels the interference, explaining why the
forward angle $\bar\nu$ neutron cross section is so small.

\begin{figure}[th]
\begin{center}
\includegraphics[scale=0.85, bb= 50 380 540 730]{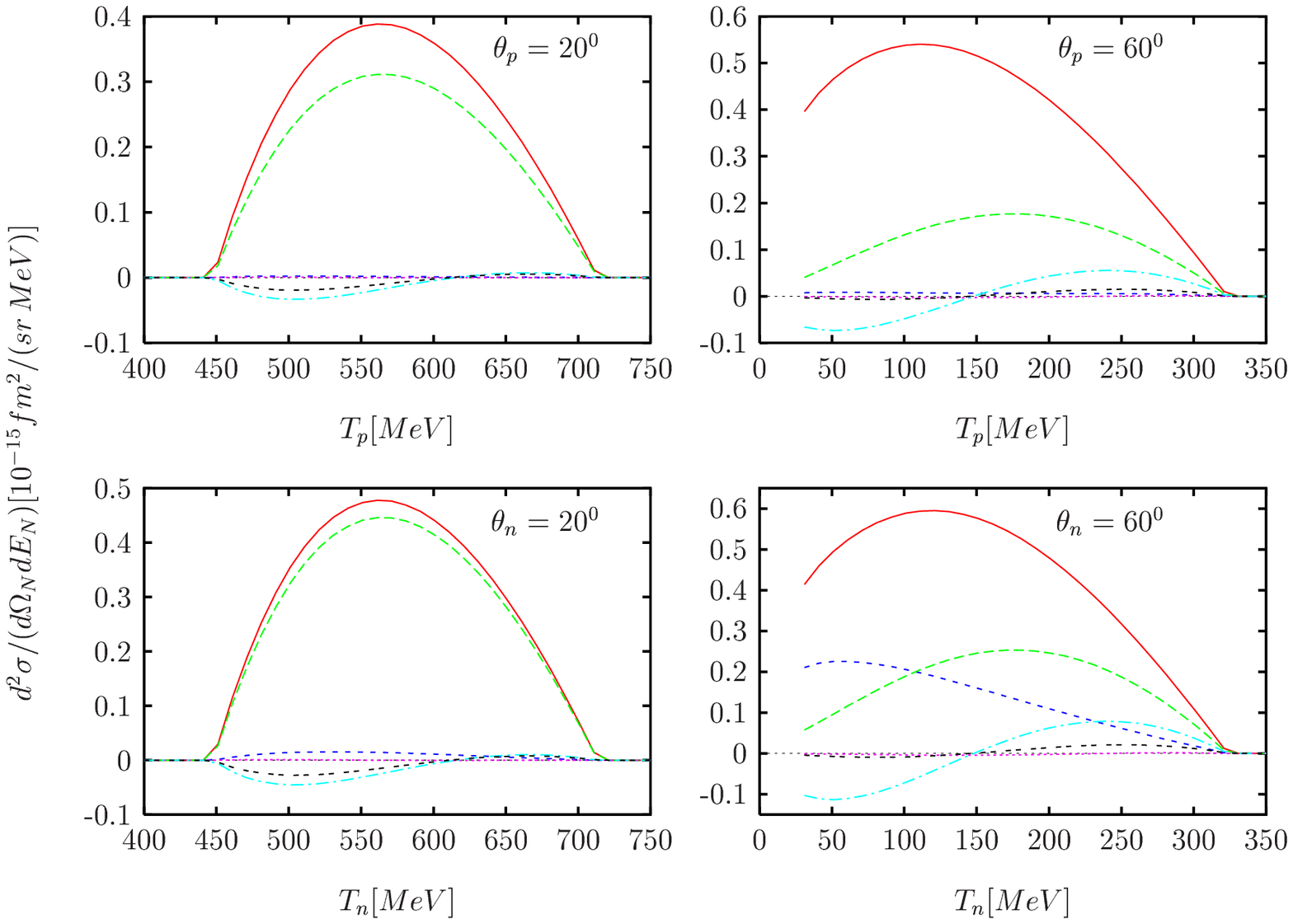}
\caption{(Color online) The separate contributions of the response
functions of Eqs.~(\protect\ref{L}-\protect\ref{TLP}) to the RFG
neutrino cross section: $L$ (short-dashed), $T$ (solid), $T'$
(dashed), ${TT}$ (dotted), ${TL}$ (dot-dashed), ${TL}'$
(double-dashed). The upper panels are for proton knockout and the
lower for neutron knockout.
\label{fig3}}
\end{center}
\end{figure}

In Fig.~\ref{fig3} the contributions of the separate responses to
the total RFG cross section are displayed. Clearly the dominant
contributions arise from the $R_T$ and $R_{T'}$ responses, and in
the case of neutron knockout, from $R_L$ at low values of the
kinetic energy. Note, however, that while not dominant (see the
discussions in the Appendix) the $R_{TL}$ response does provide an
important contribution at backward angles. In particular, since it
is negative at low kinetic energies and positive at high, it skews
the overall cross section to higher values of $T_p$ or $T_n$. Such
an effect is, as discussed above, absent for $t$-channel
scattering where the $TL$, $TL'$ and $TT$ responses are zero.

\begin{figure}[th]
\begin{center}
\includegraphics[scale=0.75, bb= 30 320 570 730]{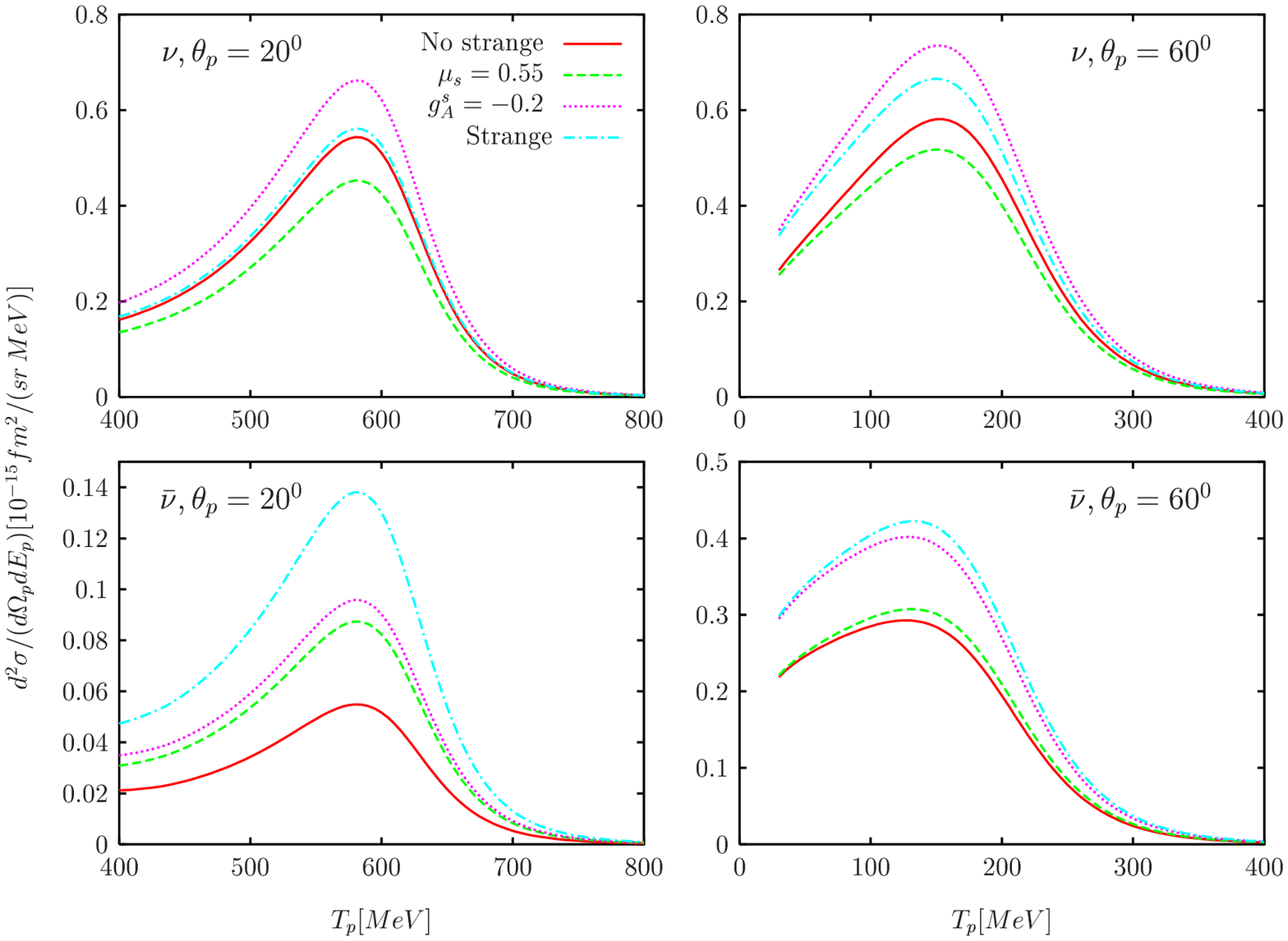}
\caption{(Color online) Effects of strangeness and radiative corrections in
neutrino and antineutrino cross sections: no strangeness (solid),
$\mu_s=0.55$ (dashed), $g_A^{s}=-0.2$ (dotted) and all the above
effects included (dot-dashed). The case of proton knockout is
assumed. \label{fig4p}}
\end{center}
\end{figure}

\begin{figure}[th]
\begin{center}
\includegraphics[scale=0.75, bb= 30 320 570 730]{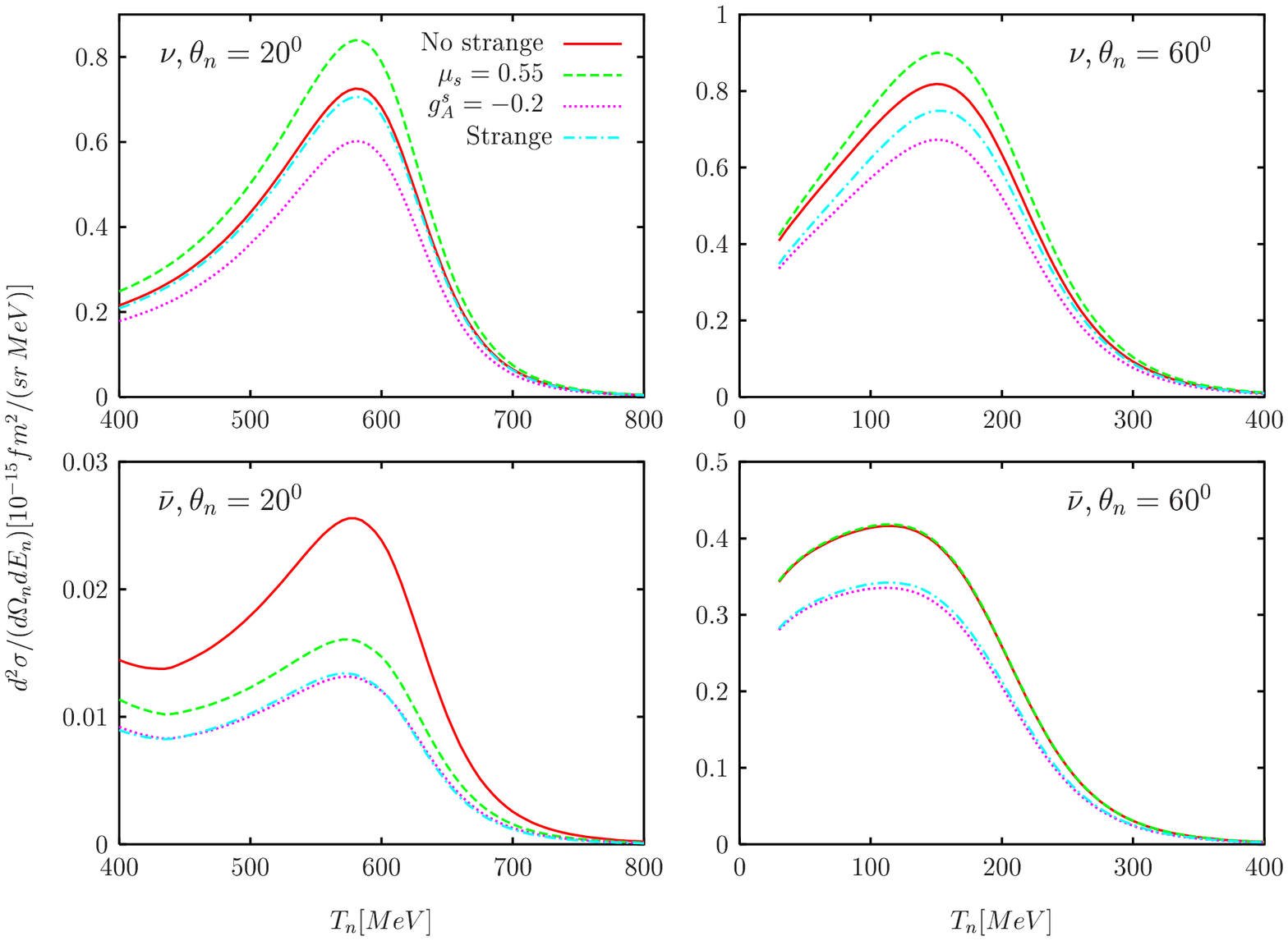}
\caption{(Color online) As for the previous figure, but now for
neutron knockout. \label{fig4n}}
\end{center}
\end{figure}

Finally, in Figs.~\ref{fig4p} (proton knockout case) and
\ref{fig4n} (neutron knockout case) we explore the dependence of
the cross section upon the strangeness content of the nucleon. We
compare the results obtained from the phenomenological
superscaling function in a situation where no strangeness is
assumed (solid line) with the ones obtained including strangeness
in the magnetic (long-dashed) and axial-vector (dotted) form
factors, using for $\mu_s=G_M^{(s)}(0)$ a representative value
extracted from the recent world studies of PV electron
scattering~\cite{strange} and taking $g_A^s=G_A^{(s)}(0)$ to be
$-0.2$~\cite{Musolf:1993tb}. The effects from inclusion of
electric strangeness are not shown here, since $G_E^{(s)}$ has
almost no influence on the full cross sections.

Starting with the proton knockout results in Fig.~\ref{fig4p}, we
see that for the $\nu$ case magnetic strangeness tends to decrease
the cross section, whereas for $\bar\nu$ it has the opposite
effect (the forward-angle $\bar\nu$ cross sections are rather
small and not considered in this discussion). For both $\nu$ and
$\bar\nu$ the axial strange contribution tends to increase the
cross section, and so the net effect of incorporating both types
of strangeness content is relatively larger in the $\bar\nu$ case
than in the $\nu$ case.

On the other hand, for the neutron knockout results shown in
Fig.~\ref{fig4n} the situation is somewhat different: for $\nu$
the roles of magnetic and axial strangeness are reversed from what
is seen for proton knockout, an effect which is easily understood
by examining the sign changes that occur in going from protons to
neutrons (see Appendix). Specifically, $G_{Mp}$ and $G_{Mn}$ are
opposite in sign, whereas, being isoscalar, $G_M^{(s)}$ is the
same for protons and neutrons. Similarly, being isoscalar
$G_A^{(s)}$ does not change sign in going from protons to
neutrons, whereas, being isovector, $G_A^{(3)}$ does. The
$\bar\nu$ case is anomalous: in this case the interference VA
response tends to cancel the VV+AA contributions. Accordingly, for
neutron knockout including magnetic strangeness, which increases
both the VV and the VA responses, has little net effect on the
cross sections, since the two effects cancel out.

\section{Conclusions}
\label{sec:sec4}

In previous work the superscaling formalism was applied to
charge-changing neutrino and antineutrino reactions with nuclei.
Using QE electron scattering data, typically at energies above
roughly 500 MeV and up to a few GeV, those analyses resulted in a
universal scaling function: scaling of both the first and second
kinds was demonstrated for the region of excitation lying below
the QE peak. This study was supplemented by an analysis of the
region lying above the QE peak where the excitation of the nucleon
to the $\Delta$(1232) dominates, and again it was shown that
scaling occurs in this domain, albeit with a different scaling
function and scaling variable, as expected. Putting these together
(two universal scaling functions and two scaling variables,
together with the elementary $eN\to eN$ and $eN\to e'\Delta$ cross
sections) one has a very good representation of inclusive electron
scattering at intermediate-to high energies from well below the QE
peak up to at least the peak of the $\Delta$-dominated region.
Importantly, this high quality agreement with experiment requires a
rather asymmetric scaling function (with a long tail extending to
high energy loss) and from other studies undertaken by us and by
others it is known that usually the results of modeling yield
nearly symmetric scaling functions, clearly at odds with the data.
It is then straightforward to insert, instead of the EM cross
sections, the elementary CC neutrino and antineutrino cross
sections to obtain CC cross sections on nuclei, as discussed in
our previous work.

In the present study we have extended that superscaling approach
now to include quasielastic scattering via the weak neutral
current of neutrinos and antineutrinos from nuclei at
intermediate-to-high energies. The same asymmetric QE scaling
function and scaling variable employed in the CC study is also
used here for the NC predictions, the essential change being
simply to insert the NC neutrino and antineutrino $\nu N$ and
${\bar{\nu}} N$ cross sections in place of the CC cross sections.
Less obvious is the application of the superscaling ideas to the
different type of inclusive reaction that must practically be
considered. Namely, while in the CC reaction studies the relevant
reaction involves an incoming lepton (a $\nu$ or ${\bar{\nu}}$)
and detection of the corresponding charged lepton at a given
scattering angle, just as in electron scattering with incident and
scattered electrons (both are $t$-channel inclusive processes),
the NC reaction is different. Here one has an incident $\nu$ or
${\bar{\nu}}$, but now a proton or neutron ejected at some angle,
whereas the scattered $\nu$ or ${\bar{\nu}}$ is not detected ---
this is a so-called $u$-channel inclusive process. Thus, in the
present work we have had to explore the validity of the
superscaling ideas when applied to such $u$-channel scattering,
again at intermediate-to-high energies where the scaling approach
can be expected to apply. Our results indicate that this is the
case and therefore that the scaling analysis used for CC reactions
should also work for NC scattering. Additionally, the use of
symmetries about the momentum transfer direction in all
unpolarized $t$-channel inclusive processes which leads to only
three independent response functions, $L$, $T$ and $T'$, is not
applicable for $u$-channel inclusive scattering. There one has in
addition the remaining responses $TL$, $TL'$ and $TT$, which
cannot be eliminated using symmetry arguments. The results
obtained in the present work show that of these only the $TL$
response appears to play a significant role, at least for the
kinematics chosen here.

In the present work scattering of neutrinos and antineutrinos at 1
GeV from $^{12}$C has been taken as representative and also since
it is relevant for on-going neutrino oscillation experiments.
Cross sections at other kinematics and for other nuclei may be
obtained by contacting the collaboration. Several conclusions
emerge from examining the results obtained.

First, the NC neutrino and antineutrino cross sections are seen to
be in roughly a 2:1 ratio ($\nu$:${\bar{\nu}}$) at backward $\nu
N$ scattering angles, whereas at forward scattering angles the
antineutrino cross sections are suppressed by an order of
magnitude or more. This holds true for both proton and neutron
knockout; moreover, the neutron knockout cross sections are
somewhat larger than the proton knockout cross sections because of
the NC single-nucleon form factors that enter in the two cases
(see text for details). These results are also rather different
from the corresponding CC reactions where it was observed that,
for the kinematics chosen, the antineutrino cross sections are
typically much smaller than for neutrinos.

Second, the interplay of the various responses ($L$, $T$, $TL$,
$TT$, $T'$ and $TL'$) is not trivial: in the various channels,
$\nu$ and ${\bar{\nu}}$, proton and neutron knockout, they play
different roles. For example, the $TL$ response is negative at low
nucleon knockout energies and positive at high energies, producing
a shift of the total cross sections to higher energies than would
occur with only the ``usual" responses $L$, $T$ and $T'$.

Finally, the effects of strangeness are relatively large and
different for the various channels, implying, as in past studies,
that high-quality measurements with $\nu$ and ${\bar{\nu}}$
together with proton and neutron knockout hold the potential to
yield more information on the strangeness content of the nucleon.

In summary, the current study employs the same superscaling
approach used previously for CC neutrino reactions now applying it
to NC neutrino scattering in the QE region. Building in the
correct scaling function, in contrast to direct modeling which
usually fails to some degree when applied to electron scattering
and therefore must surely fail to the same degree when applied to
other semi-leptonic processes, is an essential ingredient in this
approach. In on-going work our intent is to incorporate
$u$-channel inclusive cross sections for excitations in the
$\Delta$ region and beyond as in our previous work; however, such
investigations are more involved than the QE study presented here,
since then the final state involves both a nucleon and a pion, and
thus are postponed to the future.


\section*{Acknowledgments}
This work was partially supported by funds provided by Ministerio
de Educaci\'on y Ciencia (Spain) and FEDER funds, under Contracts
Nos FIS2005-01105, FPA2005-04460, FIS2005-00810, by the Junta de
Andaluc\'{\i}a, and by the INFN-MEC collaboration agreements Nos.
04-17 \& 05-22. J.A.C. also acknowledges financial support from
MEC (Spain) for a sabbatical stay at University of Torino
(PRC2005-0203). The work was also supported in part (T.W.D.) by
the U.S. Department of Energy under cooperative research agreement
No. DE-FC02-94ER40818.

\appendix*

\section{Single-nucleon cross section}
\label{sec:appA}

In this Appendix we provide the elementary cross section for the
reaction \ba \nu(\bar\nu) N &\to& \nu(\bar\nu) N ~. \label{nup}
\ea The single-nucleon cross section $\sigma\sim l_{\mu\nu}
w^{\mu\nu}$ is given in term of the leptonic tensor ( assuming
$m=m'=0$) \be l_{\mu\nu} = K_\mu K'_\nu + K'_\mu K_\nu - (K\cdot
K') g_{\mu\nu} +i\chi\epsilon_{\mu\nu\alpha\beta} K^\alpha
K'^\beta \ee with $\chi=+1$ for neutrinos and $-1$ for
antineutrinos, and of the hadronic tensor \be
w^{\mu\nu}=w^{\mu\nu}_{\cal S}+w^{\mu\nu}_{\cal A} \ . \ee This
can be decomposed into a symmetric \ba
w^{\mu\nu}_{\cal S} &=& w^{\mu\nu}_{VV}+w^{\mu\nu}_{AA}\\
&&
w^{\mu\nu}_{VV} = -w_{1V}(\tau)\left(g^{\mu\nu}+\frac{\kappa^\mu \kappa^\nu}{\tau}
\right)+w_{2V}(\tau) X^\mu X^\nu
\\
&&
w^{\mu\nu}_{AA} = -w_{1A}(\tau)\left(g^{\mu\nu}+\frac{\kappa^\mu \kappa^\nu}{\tau}
\right)+w_{2A}(\tau) X^\mu X^\nu
\nonumber\\
&&
\,\,\,\,\,\,\,\,\,\,\,\,\,\,\,\,\,\,-u_{1A}(\tau)\frac{\kappa^\mu\kappa^\nu}{\tau}
+u_{2A}(\tau)\left(\kappa^\mu\eta^\nu+\eta^\mu\kappa^\nu\right)
\ea and an antisymmetric \be w^{\mu\nu}_{\cal A} = w^{\mu\nu}_{VA}
= 2 i w_3(\tau)\epsilon^{\mu\nu\alpha\beta}
\eta_\alpha\kappa_\beta +
w_4(\tau)\left(\kappa^\mu\eta^\nu-\eta^\mu\kappa^\nu\right) \ee
tensor, where \be
X^\mu\equiv\eta^\mu+\frac{\eta\cdot\kappa}{\tau}\kappa^\mu
\stackrel{on-shell}{=}\eta^\mu+\kappa^\mu\ , \ee  having
introduced the dimensionless variables $\kappa^\mu\equiv
(\lambda,\nkappa)=Q^\mu/2 m_N$, $\eta^\mu=P^\mu/m_N$,
$\tau=\nkappa^2-\lambda^2$. Note that $u_{1A}$ (the pseudoscalar
term), $u_{2A}$ and $w_4$ do not contribute to
$l_{\mu\nu}w^{\mu\nu}$, since \be l_{\mu\nu}\kappa^\mu=(K_\mu
K'_\nu+K'_\mu K_\nu-g_{\mu\nu}K\cdot K') (K^\mu-K'^\mu)=0 \ee and
\be l_{\mu\nu}\kappa^\nu=(K_\mu K'_\nu+K'_\mu
K_\nu-g_{\mu\nu}K\cdot K') (K^\nu-K'^\nu)=0 \ee if the leptons are
massless. By contracting the above tensors we get \ba l_{\mu\nu}
w^{\mu\nu} &=& x_0\left\{ v_L R_L + v_T R_T + v_{TT} R_{TT} +
v_{TL} R_{TL} \right.
\nonumber\\
&+& \left. \chi\left(2 v_{T'} R_{T'} +2 v_{TL'} R_{TL'}\right)
\right\}\ , \ea where $x_0\equiv
2\epsilon\epsilon'\cos^2\theta_l/2$, $\theta_l$ is the lepton
scattering angle, $\rho\equiv\tau/\kappa^2$ and \ba && v_L =
\rho^2 \,\,\,\,,\,\,\,\, v_T =
\frac{1}{2}\rho+\tan^2\frac{\theta_l}{2} \,\,\,\,,\,\,\,\, v_{TT}
= -\frac{1}{2}\rho
\\
&&
v_{TL} = -\frac{1}{\sqrt{2}}\rho\sqrt{\rho+\tan^2\frac{\theta_l}{2}}
\\
&& v_{T'} =
\tan\frac{\theta_l}{2}\sqrt{\rho+\tan^2\frac{\theta_l}{2}}
\,\,\,\,,\,\,\,\, v_{TL'} =
-\frac{1}{\sqrt{2}}\rho\tan\frac{\theta_l}{2}\ . \ea The response
functions are \ba && R_L = w^{00} \label{L} \,\,\,\,,\,\,\,\, R_T
= w^{11}+w^{22}\ \label{T} ,\,\,\,\,\,\,\, R_{TT} = w^{22}-w^{11}
\label{TT}
\\
&&
R_{TL} = \sqrt{2}\left(w^{01}+w^{10}\right)
\label{TL}
\\
&& R_{T'} = i w^{21} \label{TP} \,\,\,\,,\,\,\,\, R_{TL'} =
i\sqrt{2} w^{20}\ . \label{TLP} \ea

In terms of the structure functions $w_1$, $w_2$, $w_3$ the above
response functions read (for on-shell nucleons,
$\eta\cdot\kappa=\tau$):
\ba && R_L =
-w_1(\tau)\frac{\kappa^2}{\tau}+w_2(\tau)(\varepsilon+\lambda)^2
\\
&&
R_T = 2 w_1(\tau)+w_2(\tau)\eta^2\sin^2\theta
\\
&&
R_{TT} = -w_2(\tau)\eta^2\sin^2\theta\cos(2\phi)
\\
&&
R_{TL} = 2\sqrt{2} w_2(\tau) (\varepsilon+\lambda)\eta\sin\theta\cos\phi
\\
&&
R_{T'} = 2 w_3(\tau) \frac{\tau}{\kappa}(\varepsilon+\lambda)
\\
&& R_{TL'} = 2\sqrt{2} w_3(\tau) \kappa\eta\sin\theta\cos\phi\ ,
\ea
where the angles $\theta$ and $\phi$ define the bound-nucleon
direction with respect to the reference system used in the
$t$-channel ({\bf q} along the $z$-axis), its energy being
$\varepsilon\equiv E/m_N$. Note that in this system $\phi=\phi_N$
(the outgoing nucleon's azimuthal angle).

In the usual $t$-channel inclusive scattering the $TT$, $TL$ and
$TL'$ responses vanish, since they are integrated over the
azimuthal angle $\phi$ throughout the full range $(0,2\pi)$;
however, this does not occur in $u$-channel inclusive processes,
where the integration over the outgoing lepton implies an
integration over the full range of $\phi'$, but not of $\phi$.

\mbox{}From Eqs. (A18--A23) we see that two of the responses are
proportional to the small bound-nucleon momentum parameter $\eta
\cong 1/4$; namely $R_{TL}$ and $R_{TL'}$ are both $O(\eta)$, and
therefore vanish in the limit $\eta\rightarrow 0$. Accordingly,
the $TL$ and $TL'$ responses are expected to be smaller than the
$L$, $T$ and $T'$ responses which survive in the limit
$\eta\rightarrow 0$. On the other hand, since $R_{TT}$ is
$O(\eta^2)$, one expects that the $TT$ contributions should be the
smallest, as is verified by examining the results in Sec. III.

In terms of single-nucleon form factors the structure functions
are $(a=p,n)$: \ba w_{1a}(\tau) &=& \tau \widetilde
G_{Ma}^2(\tau)+(1+\tau) \widetilde G_{Aa}^2(\tau)
\\
w_{2a}(\tau) &=& \frac{\widetilde G_{Ea}^2(\tau)+\tau \widetilde
G_{Ma}^2(\tau)}{1+\tau} + \widetilde G_{Aa}^2(\tau)
\\
w_{3a}(\tau) &=& \widetilde G_{Ma}(\tau) \widetilde G_{Aa}(\tau) \
, \ea where~\cite{Musolf:1993tb}
\begin{eqnarray}
\widetilde G_{Ep}(\tau) &=&
(2-4\sin^2\theta_W) G_E^{T=1} (\tau)
-4\sin^2\theta_W G_E^{T=0} (\tau)
- G_E^{(s) }(\tau )
\\
\widetilde G_{En} (\tau) &=& -
(2-4\sin^2\theta_W) G_E^{T=1} (\tau)
-4\sin^2\theta_W G_E^{T=0} (\tau)
- G_E^{(s)} (\tau)
\\
\widetilde G_{Mp} (\tau) &=&
(2-4\sin^2\theta_W) G_M^{T=1} (\tau)
-4\sin^2\theta_W G_M^{T=0} (\tau)
- G_M^{(s)} (\tau)
\\
\widetilde G_{Mn}(\tau)  &=& -
(2-4\sin^2\theta_W) G_M^{T=1} (\tau)
-4\sin^2\theta_W G_M^{T=0} (\tau)
- G_M^{(s)} (\tau)
\\
\widetilde G_{Ap} (\tau) &=&
-2 G_A^{(3)} (\tau) +
G_A^{(s)} (\tau)
\\
\widetilde G_{An} (\tau) &=&
2 G_A^{(3)} (\tau) +
G_A^{(s)} (\tau) ~.
\end{eqnarray}
In the above
\begin{eqnarray}
G_E^{T=0}(\tau)  &=& \frac{1}{2} \left[G_{Ep}(\tau) +G_{En}(\tau)
\right]
\\
G_E^{T=1}(\tau)  &=& \frac{1}{2} \left[G_{Ep}(\tau) -G_{En}(\tau)
\right]
\\
G_M^{T=0}(\tau)  &=& \frac{1}{2} \left[G_{Mp}(\tau) +G_{Mn}(\tau)
\right]
\\
G_M^{T=1}(\tau)  &=& \frac{1}{2} \left[G_{Mp}(\tau) -G_{Mn}(\tau)
\right]
\end{eqnarray}
 are the electromagnetic isoscalar and isovector Sachs form factors,
 whereas the isovector axial-vector form factor is given by
\be G_A^{(3)}(\tau)  = \frac{1}{2} (D+F) G_A^D(\tau) \ee with
$D=1.262/1.64$, $F=0.64 D$ and $G_A^D(\tau) =(1+3.32\tau)^{-2}$.

The strangeness form factors are parameterized as follows:
\begin{eqnarray}
G_E^{(s)}(\tau)  &=& \rho_s\tau G_V^D(\tau)
\\
G_M^{(s)}(\tau)  &=& \mu_s G_V^D(\tau)
\\
G_S^{(s)}(\tau)  &=& g_A^s G_A^D(\tau) ~,
\end{eqnarray}
with $G_V^D(\tau)=(1+4.97\tau)^{-2}$.

\end{document}